\documentclass{article}
\usepackage{color}
\usepackage{graphicx}
\usepackage{subfig}
\usepackage{url}
\usepackage{amsmath}
\usepackage{float}
\usepackage{wrapfig}
\usepackage{url}
\usepackage{hyperref}
\usepackage{xspace}
\usepackage[disable]{todonotes}
\usepackage[plain]{algorithm}

\usepackage{tikz}
\usepackage{pgfplots}
\usepackage{ifthen}

\newcommand{\TODO}[1]{}

\usepackage{etoolbox}
\makeatletter
\preto{\@verbatim}{\topsep=0pt \partopsep=0pt }
\makeatother

\usepackage{listings}
\usepackage{color}

\definecolor{dkgreen}{rgb}{0.5,0.5,0.5}
\definecolor{gray}{rgb}{0.5,0.5,0.5}
\definecolor{mauve}{rgb}{0.58,0,0.82}
\definecolor{lightgray}{rgb}{0.93,0.93,0.98}
\definecolor{keyw}{rgb}{0.2,0.2,1}

\lstdefinelanguage{scala}{
  morekeywords={abstract,case,catch,class,def,%
    do,else,extends,false,final,finally,%
    for,if,implicit,import,match,mixin,%
    new,null,object,override,package,%
    private,protected,requires,return,sealed,%
    super,this,throw,trait,true,try,%
    type,val,var,while,with,yield},
  otherkeywords={=>,<-,<\%,<:,>:,\#,@},
  sensitive=true,
  morecomment=[l]{//},
  morecomment=[n]{/*}{*/},
  morestring=[b]",
  morestring=[b]',
  morestring=[b]"""
}

\newcommand{\ignore}[1]{}

\newcommand{\framework}{{\rmfamily\scshape FooPar}\xspace}

\title{\framework: A Functional Object Oriented Parallel Framework in Scala}

\author{Felix P. Hargreaves and Daniel Merkle\\
Department of Mathematics and Computer Science\\ University of Southern Denmark\\\texttt{\{daniel,felhar07\}@imada.sdu.dk} }
\date{11. June 2013}

\begin{document}

%

\maketitle

\begin{abstract}
  We present FooPar, an extension for highly efficient Parallel
  Computing in the multi-paradigm programming language Scala. Scala
  offers concise and clean syntax and integrates functional
  programming features. Our framework FooPar combines these features
  with parallel computing techniques. FooPar is designed modular and
  supports easy access to different communication backends for
  distributed memory architectures as well as high performance math
  libraries. In this article we use it to parallelize matrix-matrix
  multiplication and show its scalability by a isoefficiency
  analysis. In addition, results based on a empirical analysis on two
  supercomputers are given. We achieve close-to-optimal performance
  wrt. theoretical peak performance.  Based on this result we conclude
  that FooPar allows to fully access Scala’s design features without
  suffering from performance drops when compared to implementations
  purely based on C and MPI.
\end{abstract}
\TODO{
a reminder TODO:
\begin{itemize}
\item consistent writing: zip, map, reduce (which font?)
\item parallel system/algorithm (which phrase?)
\item computing elemnts, processors, ... (what to chose?)
\item Scientific References! Anything below 15 to 20 is a no-go.
\item don't use ``prototype implementation''. We use ``generic''.
\end{itemize}
}

	\addtolength{\abovedisplayskip}{-2pt}
	\addtolength{\belowdisplayskip}{-2pt}
	\addtolength{\abovecaptionskip}{-2pt}
        \addtolength{\belowcaptionskip}{-2pt}
	\addtolength{\abovedisplayshortskip}{-2pt}
	\addtolength{\belowdisplayshortskip}{-2pt}
        \addtolength{\textfloatsep}{-2pt}
\section{Introduction}
\todo[inline]{after reading our old intro, I actually can understand the reviewers critisism. we need a more general intro (1-3 paragraphs). and something like: to the best of our knowledge, there is no other functional programming based framework that reaches  ... or something similar, we should make more explicitly clear that to the best of our knowledge our framework is unique. there are to many ``highly efficient'', and positive phrases in the intro, it's too FooPar focussed, it sounds too much like ``selling an a non-objective manner''. we need to give it a more critical bias, and it should be clear why we can not or don't need to compare to other frameworks. I already removed some phrases, please be self-critcial.\\[1cm]

Intro suggestion: functional programming and functional programming concepts are becoming more and more relevant ... examples include ... lambda expressions in C++ standard (check! it is in g++ 4.8 now ...) extension, armada of companies behind scala. a usually used arument against is performance issues, especially if it comes to HPC. With foopar we aim at bridging this gap.  }

Functional programming is becoming more and more ubiquitous (lambda functions introduced in C++11 and Java8) due to higher levels of abstraction, better encapsulation of mutable state, 
and a generally less error prone programming paradigm. In HPC settings, the usual argument against the added functional abstraction is performance issues. \framework aims to
bridge the gap between HPC and functional programming by hitting a sweet spot between abstraction and efficiency not addressed by other functional frameworks.

There exists a multitude of map-reduce based frameworks similar to Hadoop which focus on big data processing jobs, often in cloud settings. 
Other functional parallel frameworks like  Haskell's \textit{Eden} \cite{eden05} and Scala's \textit{Spark} \cite{spark10} focus on workload balancing strategies neglecting performance to increase abstraction.
While many different functional frameworks are available, most seem to value abstraction above all else. 
To the best of our knowledge, other functional parallel frameworks do not reach asymptotic or practical performance goals comparable to \framework.

In this paper (after definitions and a brief introduction to isoefficiency in Section~\ref{sec:def}) we will introduce \framework in Section~\ref{secFramework} and describe its
architecture, data structures, and 
operations it contains. The complexity of the individual operations on the (parallel) data structures will be shown to serve as basis for parallel complexity analysis.
A matrix-matrix multiplication algorithm will be designed using the
functionality of \framework; the implementation will be analyzed with an
isoefficiency analysis in Section~\ref{sec:mat}. Test
results showing that \framework can reach close-to theoretical peak performance
on large supercomputers 
will be presented in Section~\ref{testresults}. We conclude with Section~\ref{sec:con}.

\section{Definitions, Notations, and Isoefficiency}
\label{sec:def}
The most widespread model for scalability analysis of heterogeneous
parallel systems (i.e. the parallel algorithm and the parallel
architecture) is isoefficiency \cite{Kumar87}\cite{Grama93} analysis. The
{\em isoefficiency function} for a parallel system relates the {\em
  problem size} $W$ and the {\em number of processors} $p$ and defines
how large the problem size as a function in $p$ has to grow in order
to achieve a constant pre-given efficiency. Isoefficiency has been
applied to a wide range of parallel systems (see, e.g.
\cite{dns},\cite{Hwang98},\cite{hiso}). As usual, we will define the {\em
 message passing costs}, $t_c$, for parallel machines as $t_c := t_s + t_w
\cdot m$, where
$t_s$ is the start-up time, $t_w$ is the per-word transfer time, and $m$
is the message size. The {\em sequential (resp. parallel) runtime}
will be denoted as $T_S$ (resp. $T_P$). The {\em problem size} $W$ is
identical to the sequential runtime, i.e. $W:=T_S$. The {\em overhead function}
will be defined as $T_o(W,p) := p T_P - T_S$ .
The
isoefficiency function for a parallel system is usually found by an
algebraic reformulation of the equation $W = k \cdot T_o(W,p)$ such
that $W$ is a function in $p$ only (see e.g. \cite{Grama93} for more
details). In this paper we will employ broadcast and reduction
operations for isoefficiency analysis for parallel matrix-matrix
multiplication with \framework. Assuming a constant cross-section
bandwith of the underlying network and employing recursive doubling
leads to a one-to-all broadcast computational runtime of $(t_s + t_w
\cdot m) \log p$ and the identical runtime for an all-to-one
reduction with any associative operation $\lambda$. All-to-all
broadcast and reduction have a runtime of $t_s \log p + t_w \cdot (p -
1)$. A circular shift can be done in runtime $t_s +
t_w\cdot m$ if the underlying network has a cross-section bandwith of
$O(p)$.

A parallel system is {\em cost-optimal} if the processor-time product has
the same asymptotic growth as the parallel algorithm, i.e.
$p\cdot T_P \in \Theta(T_S)$.


\section{The \framework Framework}
\label{secFramework}
\framework is a modular extension to Scala\cite{pis} which supports user extensions and additions to data structures with proven Scala design patterns. 
Scala is a \textit{scalable language} pointing towards its ability to
make user defined abstractions seem like first class citizens in the language.
The object oriented aspect leads to concise and readable syntax when combined
with operator overloading, e.g. in matrix operations. Scala unifies functional and imperative programming making it ideal for high performance computing. It builds on the Java Virtual Machine (JVM) which is a mature
platform available for all relevant architectures. Scala is completely interoperable with Java, which is one of the reasons why many companies move their performance critical code to a Scala code base \cite{enter}.
Today, efficiency of byte-code can approach that of optimized C-implementations within
small constants \cite{comp}. Further performance boosts can be gained by using \textit{Java Native Interface}; 
however, this adds an additional linear amount of work due to memory being
copied between the virtual
machine and the native program. In other words, \textit{super linear workloads} motivate the usage of \textit{JNI}.

\begin{figure}[t]
 \centering
 \includegraphics[scale=0.5]{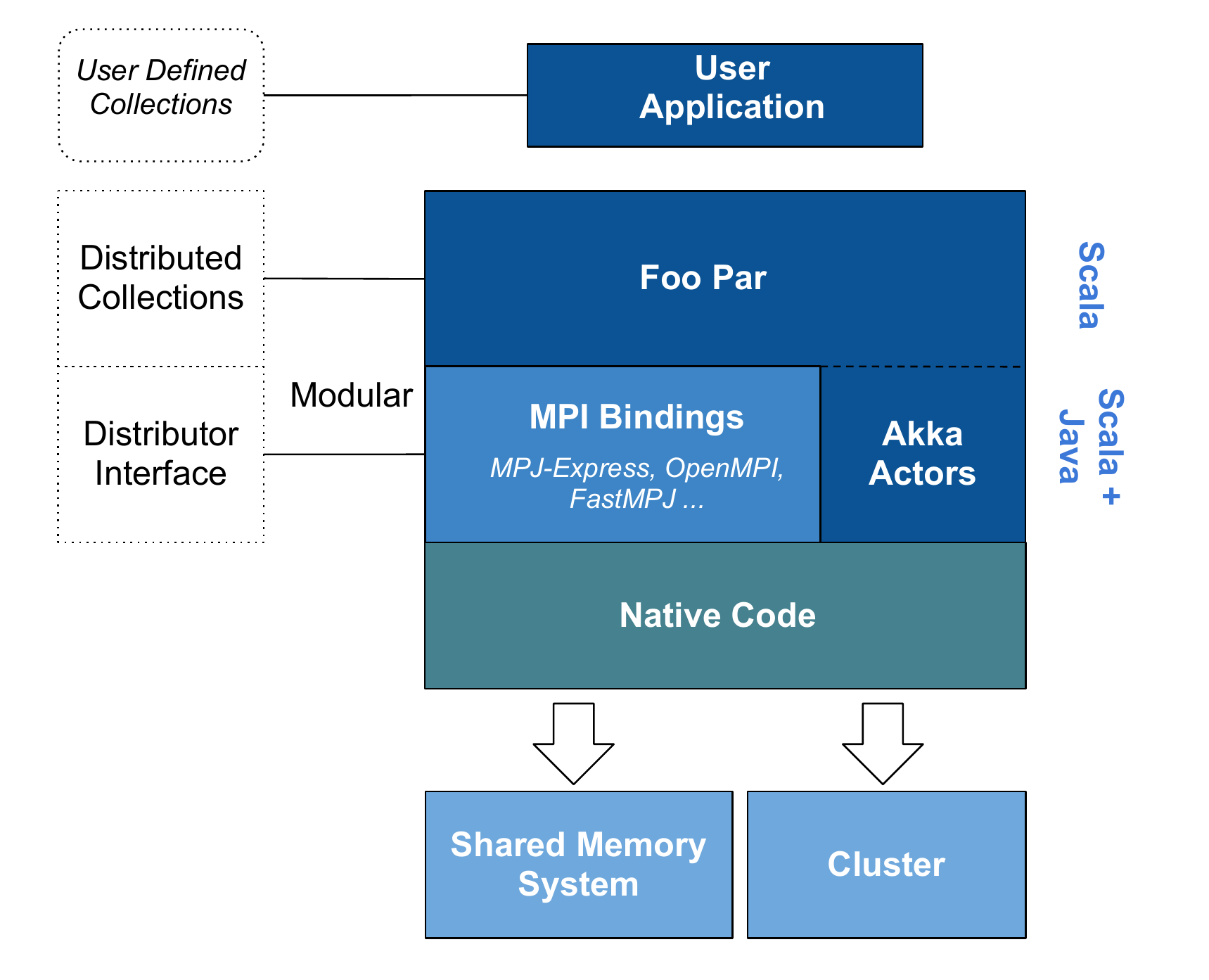}
 \caption{Conceptional overview of the layered architecture of \framework. \TODO{expand with color boxes/subscript notation definition as discussed }}
 \label{architecture}
\end{figure}

Fig.~\ref{architecture} depicts the architecture of \framework.
Using the \textit{builder/traversable} pattern \cite{fsttcs2009}, one can create
maintainable distributed collection classes while benefiting from the underlying
modular communication layer.
In turn, this means that user provided data structures receive the same benefits from the remaining layers of the framework as the ones that ship with \framework.
It is possible to design a large range of parallel algorithms using purely the data structures within \framework although one is nor restricted to that approach.

A configuration of \framework can be described as \framework-X-Y-Z, where X is
the communication module, and Y is the native code used for networking and Z
is the hardware configuration, e.g. X$\in \{$
MPJ-Express, OpenMPI, FastMPJ, SharedMemory, Akka $\}$, Y$\in \left\{
\text{MPI, Sockets} \right\}$ and Z$\in \left\{ \text{SharedMemory, Cluster,
Cloud} \right\}$. Note that this is not an exhaustive listing of module
possibilities. In this paper
we only use Y=MPI and Z=Cluster and do not analyze Shared Memory
parallelisation. Therefore, we will only use the notation \framework-X.




\subsection{Technologies}

Currently, \framework uses the newest version of Scala 2.10. The \textit{Scalacheck} framework is a specification testing framework for Scala which is used to test the methods provided by \framework data structures.
\textit{JBLAS}, a high performing linear algebra library \cite{jmat} using BLAS via JNI is used
to benchmark \framework with an implementation of distributed matrix-matrix
multiplication. Intel\textsuperscript{\textregistered}'s  {\em Math Kernel
Library } offers an high-performing alternative with Java bindings, and will
also be used for benchmarking.

\subsection{SPMD Operations on Distributed Sequences}
\TODO{split into concept / example}
\framework is inspired by the \textit{SPMD/SIMD Principle} often seen in
parallel
hardware \cite{spmd}. \TODO{refs} \TODO{...but, can't find articles about maybe
monad. Should we put something just about monads? It is probably more
beneficial!}
The \textit{Option} monad in \textit{Scala} is a construct similar to \textit{Haskell's} \textit{maybe monad}.
Option is especially suited for SPMD patterns since it supports \texttt{map} and
\texttt{foreach} operations.
Listing \ref{list:spmd} exemplifies this characteristic approach in \framework.
\begin{lstlisting}[float=ht,caption=SPMD example,label=list:spmd]
def ones(i: Int): Int = i.toBinaryString.count(_ == '1')
val seq = 0 to worldSize - 3
val counts = seq mapD ones
println(globalRank+":"+counts)
\end{lstlisting}
Here, \texttt{ones(i)} counts the number of 1's in the binary representation of
\texttt{i}.\: \texttt{mapD}~distributes the map operation on the Scala range
\texttt{seq}.

In SPMD, every process runs the same program, i.e. every process generates \texttt{seq}
in line 3. If combined with lazy-data objects, this does not lead to unnecessary space or
complexity overhead (cmp. Fig.~2 and~3). 
While every process generates the sequence, only \textit{some} processes perform the \texttt{mapD} operation.

\TODO{Show output and code, side by side.}
\begin{figure}[t]
\centering
\begin{minipage}{.5\textwidth}
  \centering
\begin{tikzpicture}
    
    \def \s {0.43}
    \def \x {4}
    \def \n {2}
    \pgfmathtruncatemacro{\pn}{\n+2}
    \pgfmathtruncatemacro{\nplus}{\n+1}
    \pgfmathsetmacro{\shalf}{\s*0.5}

    \node at (\x-1,0.3) {$Ranks$};
    \node at (\x+\shalf,0.3) {$seq$};
    \node at (\x+\s+\s+0.9,0.3) {$Operation$};

    \foreach \y in { 0,...,\n} 
    {
      \draw [fill=lightgray,ultra thick] (\x,-\y*\s) rectangle
(\x+1*\s,-\y*\s-\s);
      \node at (\x+0.5*\s,-\y*\s-0.5*\s) {\y};
      
      \draw [->,thick] (\x-\s,-\y*\s-\shalf) -- (\x,-\y*\s-\shalf);
      \draw [->,thick] (\x+\s,-\y*\s-\shalf) -- (\x+\s+\s,-\y*\s-\shalf);
      \node at (\x+\s+\s+0.9,-\y*\s-\shalf) {$\lambda(seq_\y)$};
    }
    
    \foreach \y in { 0,...,\pn} 
    {
      \node at (\x-1,-\y*\s-0.5*\s) {$p_\y$};
    }
    
    \foreach \y in { \nplus,...,\pn} 
    {
      \draw [->,thick] (\x-\s,-\y*\s-\shalf) -- (\x+\s+\s,-\y*\s-\shalf);
      \node at (\x+\s+\s+0.9,-\y*\s-\shalf) {\textit{nop}};
    }
    \end{tikzpicture}
  \label{fig:test1}
    \captionof{figure}{A distributed map operation.}
\end{minipage}%
\begin{minipage}{.5\textwidth}
  \centering

  \begin{verbatim}


       3:DSeq(None)
       2:DSeq(Some(1))
       0:DSeq(Some(0))
       1:DSeq(Some(1))
       4:DSeq(None)
  \end{verbatim}
  \label{fig:test2}
    \captionof{figure}{Output of the distributed map operation (arbitrary order).}
\end{minipage}

\end{figure}

 \vspace{-0.4cm}
\subsection{Data Structures}
\vspace{-0.2cm}
\TODO{move after SPMD}
\framework relies heavily on the interpretation of data structures as \textit{process-data} mappings. As opposed to many modern parallel programming tools, \framework 
uses static mappings defined by the data structures and relies on the user to
partition input. This decision was made to ensure efficiency and analyzability.
By using static mappings in conjunction with {\em SPMD}, the overhead and
bottleneck pitfalls induced by {\em master slave} models are avoided and
program-simplicity and efficiency are achieved.
In \framework, data partitioning is achieved through \textit{proxy-} or
\textit{lazy} objects, which are easily defined in \textit{Scala}.
In its current state, \framework supports distributed singletons (aka.
distributed variables), distributed sequences and distributed multidimensional
sequences.
The distributed sequence combines the notion of \textit{communication groups}
and data. 
By allowing the dynamic creation of communication groups for sequences, a total abstraction of network communication is achieved. 
Furthermore, a communication group follows data structures for subsequent
operations allowing for advanced chained functional programming to be highly
parallelized.  
Tab.~\ref{dseq} lists a selection of supported operations on distributed
sequences. The given runtimes are actually achieved in \framework, but of course they depend on the implementation of collective operations in the communication backend.
A great advantage of excluding user defined message passing is gaining analyzability through the provided data-structures. 

\begin{table}[t]
\hspace*{-0.7cm}

\begin{tabular}{|l|p{3.6cm}|p{3.2cm}|c|}\hline
 \textbf{Operation} & \textbf{Semantic} & \textbf{Notes} & \textbf{$T_p$ (parallel runtime)}\\\hline\hline
 \texttt{mapD($\lambda$)} & \textit{Each process} transforms one element of the
sequence using operation~$\lambda$  (element size $m$)  & This is a
non-communicating operation  & $\Theta(T_\lambda(m))$\\\hline
 \texttt{reduceD($\lambda$)} & The sequence with $p$ elements is reduced to the \textit{root process} using operation~$\lambda$& $\lambda$ must
be an \textit{associative} operator& $\Theta( \log p (t_s + t_w m
+T_\lambda(m))) $\\\hline
 \texttt{allGatherD} & All processes obtain a list where element $i$ comes from
process $i$ & Process $i$ provides \textit{the valid} $i$th element& $\Theta(
(t_s +t_w m) (p-1) )$\\\hline
 \texttt{apply(i)} & All processes obtain the $i$th element of the sequence & sementically identical to a one-to-all broadcats
& $\Theta( \log p (t_s+t_w m))$\\\hline
\end{tabular} 
\caption{A selection of operations on distributed sequences in \framework. }
%
\label{dseq}
\end{table}

\section{Matrix-Matrix Multiplication in \framework}
\label{sec:mat}
\TODO{sec 3.1 : sequential, 3.2: parallel. similarity to DNS (which comes as a result)}

\subsection{Serial Matrix-Matrix Multiplication}

Due to the abstraction level provided by the framework, algorithms can be defined in a fashion which is often very similar to a mathematical definition. 
Matrix-matrix multiplication is a good example of this. The problem can be defined as follows:
\begin{align*}
 (AB)_{i,j} := \sum_{k=0}^{n-1} A_{i,k} B_{k,j}
\end{align*}
\noindent
where $n$ is the number of rows and columns in matrices $A$ and $B$ respectively. \TODO{Felix: why M and not just n as used later?}
In functional programming, list-operations can be used to model this expression in a concise manner.
The three methods, \textit{zip, map} and \textit{reduce} are enough to express
matrix-matrix multiplication as a functional program. A serial algorithm for
matrix-matrix multiplication based on a 2d-decomposition of the matrices could look like this:
\begin{align*}
C_{i,j} &\leftarrow reduce \; ( + )  \; (zipWith \; ( \cdot)  \; A_{i*} \; B^T_{*j}) ,  & & \forall (i,j) \in   \mathcal{R} \times \mathcal{R}
\end{align*}
Here, $\mathcal{R} = \{0,\ldots,q-1\}$, and the sub-matrices are of size $(n/q)^2$. Operation \textit{zipWith} is a convenience method roughly equivalent to: $map \circ zip$,  which takes 2 lists and a 2-arity function to combine them. 

\subsection{Generic Algorithm for Parallel Matrix-Matrix Multiplication}

To illustrate the simplicity of complexity analysis, the parallel version of the
algorithm can be written in a more verbose fashion as follows:
\begin{align*}
C_{i,j} \leftarrow \texttt{reduceD} \; ( + )  \;(\texttt{mapD} \; (\cdot )\;
(\texttt{zip} \; A_{i*}
\; B^T_{*j})) ,  & &\forall (i,j) \in   \mathcal{R} \times \mathcal{R}
\end{align*}
Operation \texttt{zip} is $\in \Theta(1)$ due to lazy
evaluation. We use a block size $m=(n/q)^2$. For \texttt{mapD} (multiplication of
sub-matrices) we have $T_{\hbox{\tiny mult}}(m)=\Theta(m^{3/2})$, for \texttt{reduceD}
(summation of sub-matrices) we have $T_{\hbox{\tiny sum}}(m)=\Theta(m)$. In
asymptotic terms the \textit{parallel runtime} $T_P$ is therefore: 
$$ T_P = \overbrace{\Theta(1)}^{\texttt{zip}} +
\overbrace{\Theta((n/q)^3)}^{\texttt{mapD}} + \overbrace{\Theta((n/q)^2 \log
q)}^{\texttt{reduceD}} $$
Since $C_{i,j}$ is independent both in $i$ and $j$, the $q^2$ operations can all run in parallel. Using $q$ processors per reduction leads then to $p = q^2 \cdot q$ processors and the overall asymptotic runtime  $\Theta((n/p)^2 \log p)$.

Using the framework, some parts of the analysis can be carried out independently
of the \textit{lambda operations} used in an algorithm. 
What is left is a \textit{generic algorithm} which shows precisely the communication pattern used in the algorithm. As a coincidence, the communication pattern is essentially 
identical to that of the well known DNS algorithm~\cite{dns81},\cite{dns}.

Algorithm \ref{alg:protomult} shows a complete \framework implementation, which
is practically identical to the pseudo code. Note, that the algorithm uses
proxy-objects which
are simply objects containing lazy data using Scala's \textit{lazy} construct~\cite{scalref}.
\begin{algorithm}[t]
\begin{lstlisting}
//Initialize matrices
val A = Array.fill(M, M)(MJBLProxy(SEED, b))
val Bt = Array.fill(M, M)(MJBLProxy(SEED, b)).transpose

//Multiply matrices
for (i <- 0 until M; j <- 0 until N)
  A(i) zip Bt(j) mapD { case (a, b) => a * b } reduceD (_ + _)
\end{lstlisting}
 \caption{Generic algorithm for matrix-matrix multiplication with \framework.}
 \label{alg:protomult}
 \end{algorithm}

\subsubsection{Isoefficiency Analysis for the Generic Algorithm:}
\label{sec:iso}
\TODO{move code to sec 3. Merge 4.1 and 4.2, Intro: DNS has isoefficiency of blabla ... in 4.1 we will present a straight-forward method (more details!) bla bla ... in 4.2 we rule the world ... do results on 4.1 and 4.2 }
We start by determining the non-asymptotic parallel runtime. We assume
the number of processors is $p=q^3$ (i.e. $q=p^{1/3}$) and matrices
$A$ and $B$ of size $n\times n$. Splitting $A$ and $B$ into $q\times
q$ blocks leads to a block size of $(n/q)^2$. The \texttt{zip}
operation has a runtime of $q^2$ due to \texttt{nop} instructions
carried out in iterations where the current process is not assigned to
the operation. An implicit conversion (runtime $q^2$) is needed to
extend the functionality of standard Scala arrays. The \texttt{mapD}
operation has a runtime of $q^2+(n/q)^3$ and the \texttt{reduceD} operation has a
runtime of $q^2 + \log q + (n/q)^2 \log q$. As $q^2 = p^{2/3}$, this
leads to an overall parallel runtime of $$ T_p = 4\cdot p^{2/3}
+\frac{n^3}{p} +1/3 \left( \log p + \left(\frac{n^2}{p^{2/3}}\right)
  \log p \right),$$ and the corresponding cost $p\cdot T_P \in \Theta(
4p^{5/3}+n^3)$. Therefore this approach is cost-optimal for $p \in
O(n^{9/5})$. The overhead for this basic implementation is $$T_o =
pT_p - T_S = 4p^{5/3} +\frac{p}{3} \left( \log p +
  \left(\frac{n^2}{p^{2/3}}\right) \log p \right).$$ Following an
isoefficiency analysis based on $W = K\cdot T_o(W,p)$ leads to
$$ W = n^3 = K 4p^{5/3} +K p\left( \log p + \left(\frac{n^2}{p^{2/3}}\right)
\log p \right).$$
Examining the terms individually shows that the first term of $K\cdot
T_o(W,p)$ constraints the scalability the most. Therefore, the
isoefficiency function for the basic algorithm is $W \in \Theta(p^{5/3})$.
Fig.~\ref{fig:3d} shows the communication pattern implemented by Algorithm
\ref{alg:protomult}.

\subsection{Grid Abstraction in \framework for Parallel Matrix-Matrix Multiplication}

In \cite{grama} an isoefficiency function in the order of  $\Theta( p \log^3 p )$ was achieved by using the DNS algorithm for matrix-matrix multiplication. 
The bottleneck encountered in the basic implementation is due to the inherently
sequential \texttt{for loop} emulating the $\forall$ quantifier. 
Though Scala offers a lot of support for \textit{library-as-DSL} like patterns,
there is no clear way to offer safe parallelisation of nested for loops while
still supporting 
distributed operations on data structures. 
To combat this problem, \framework supports multidimensional distributed
sequences in conjunction with constructors for arbitrary Cartesian grids.
\texttt{Grid3D} is a special case of \texttt{GridN}, which supports iterating
over 3D-tuples as opposed to \textit{coordinate lists}. 
Using \texttt{Grid3D} an algorithm for matrix-matrix multiplication can be implemented
as seen in Algorithm \ref{alg:gridmult}.
\begin{figure}[t]
 \centering
 \includegraphics[scale=0.3]{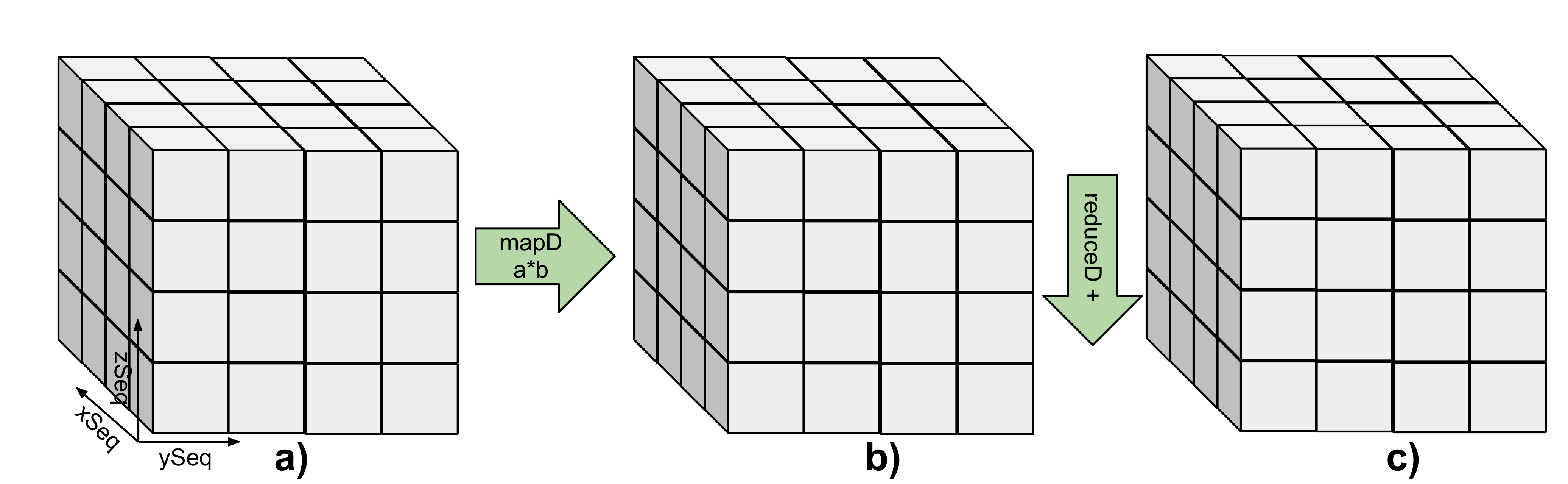}
 \caption{\textbf{a)} Process $(i,j,k)$ contains blocks $A_{i,k}$ and $B_{k,j}$
\textbf{b)}
local multiplication $C_{i,j} = A_{i,k}\times B_{k,j}$, \textbf{c)} after reduction (summation): process $(i,j,0)$ contains the (partial) result matrix.}
\label{fig:3d}
\end{figure}
\begin{algorithm}[t]
\begin{lstlisting}[language=Scala,tabsize=3,frame=single]
val R = 0 until DIM
val G = Grid3D(R, R, R)

val GA = G mapD { case (i, j, k) => A(i)(k) }
val GB = G mapD { case (i, j, k) => B(k)(j) }

val C = ((GA zipWithD GB)(_ * _) zSeq) reduceD (_ + _)
\end{lstlisting}
 \caption{Matrix-matrix multiplication in \framework using Grid Abstraction.}
 \label{alg:gridmult}
\end{algorithm}
\texttt{zSeq} is a convenience method for getting the distributed
sequence, which is variable in $z$ and constant in the $x,y$
coordinates of the current process. By using the grid data structure,
we safely eliminate the
overhead induced by the for-loop in Algorithm \ref{alg:protomult} and end up
with the same basic communication pattern as shown in Fig.~\ref{fig:3d}.
Operation
\texttt{mapD} has a runtime of $\Theta((n/q)^3)$ and \texttt{reduceD} a runtime
of
$\Theta(\log q + (n/q)^2 \log p)$. Due to space
limitations we will not present the details of runtime and isoefficiency
analysis but refer to \cite{dns}, as the analysis given there is 
very similar. Parallel runtime, $T_P$, and cost are given by
  $T_P = n^3/p + \log p + \left(n^2/p^{2/3}\right) \log p$ and
  $\hbox{cost} \in \Theta(n^3 + p \log p + n^2 p^{1/3} \log p)$. This leads to an isoefficiency function in the order of $\Theta( p \log^3 p )$, identical to the isoefficiency achieved by the DNS algorithm.






\TODO{DANIEL, can you please try to find the article describing the isoefficiency analysis for DNS?}

\section{Test Results}
\label{testresults}
~\\{\noindent \bf Parallel Systems and their Interconnection
  Framework:} In this study we focus on analyzing scalability,
efficiency and flexibility. We tested \framework on two parallel
systems: the first system is called Carver and is used to analyze the
{\em peak performance} and the overhead of \framework. It is an IBM
iDataPlex system where each computing node consists of two Intel Nehalem
quad-core processors (2.67 GHz processors, each node has at least 24GB
of RAM). The system is located at the Department of Energy's National
Energy Research Scientific Computing Center (NERSC).  All nodes are
interconnected by 4X QDR InfiniBand technology, providing maximally 32
Gb/s of point-to-point bandwidth. A highly optimized version of
Intel's Math Kernel Library (MKL) is used, which provides an empirical
peak performance of $10.11$ GFlop/s on one core (based on a single core
matrix-matrix multiplication in C using MKL). This will be our reference performance
to determine efficiency on Carver. Note, that the empirical peak performance is
very close to the theoretical peak performance of $10.67$ GFlop/s on
one node. The largest parallel job in Carver's queuing system
can use maximally 512 cores, i.e. the theoretical peak is 5.46 TFlop/s.

The second system has basically the same hardware setup. The name of
the system is Horseshoe-6 and it is located at the University of
Southern Denmark. Horseshoe-6 is used in order to test the {\em
  flexibility} of \framework. The math libraries are not compiled
towards the node's architecture, but a standard high performing BLAS
library was employed for linear algebraic operations. The reference
performance on one core was measured again by a matrix-matrix
multiplication (C-version using BLAS) and is 4.55 GFlop/s per core.

On Carver Java bindings of the nightly-build OpenMPI version
1.9a1r27897 \cite{gabriel04} were used in order to
interface to OpenMPI (these Java bindings are not yet
available in the stable version of OpenMPI). On Horseshoe-6 we used
three different communication backends, namely i.) OpenMPI Java
bindings (same version as on Carver), ii.) MPJ-Express
\cite{shafi09}, and iii.)  FastMPJ
\cite{taboada_jos10}. Note, that changing the communication backend does not
require any change in the Scala source code for the parallel
algorithmic development within \framework.

For performance comparison of \framework and C we also developed a highly
optimized parallel version of the DNS algorithm for matrix-matrix
multiplication, using C/MPI. MKL (resp. BLAS) was used on Carver (resp.
Horseshoe-6) for the sub-matrix-matrix multiplication on the
individual cores. Note, that the given efficiency results basically do not
suffer any noticable fluctuations when repeated.

\begin{figure}[t]
\begin{minipage}[b]{.5\textwidth}
\begin{tikzpicture}[scale=0.75]
\begin{axis}[
height=6cm,
width=8.6cm,
legend style={
cells={anchor=east},
at={(0.5,1.55)},anchor=north,
legend plot pos=right
},
xlabel={$p$},
legend columns=3,
ymax=100, ymin=0,
xmin=8, xmax=512,
grid=major,
symbolic x coords={8,27,64,125,216, 343, 512},
cycle multi list={
color list\nextlist
[3 of]mark list
}]]

\addplot coordinates {
(216, 93.5450587056656)
(343, 90.7414150182952)
(512, 85.3878912543401)
}; \addlegendentry{\framework-OpenMPI: 30240}

\addplot coordinates {
(27, 91.9443745666318)
(64,89.8422222804702)
(125, 85.3684289707806)
(216, 66.0358723251896)
}; \addlegendentry{12600}

\addplot coordinates {
(8,83.6735621581713)
(27,54.4758084131078)
(64, 58.0639775889845)
(125, 41.7656664437993)
(216, 8.92772579289919)
}; \addlegendentry{4200}

\addplot coordinates {
(216, 94.7927932432726)
(343, 92.3447717731622)
(512, 92.8924305968514)
}; \addlegendentry{C-MPI: 30240}

\addplot coordinates {
(27, 90.2057518951318)
(64, 93.5781686991501)
(125, 88.6933351263216)
(216, 64.5340563655128)
}; \addlegendentry{ 12600}

\addplot coordinates {
(8, 97.3957918015619)
(27, 86.0607170894597)
(64, 72.4357326986336)
(125, 43.3537528654884)
(216, 21.8396035237039)
}; \addlegendentry{4200}

\end{axis}
\end{tikzpicture}
\end{minipage}
\begin{minipage}[b]{.5\textwidth}
\hfill\begin{tikzpicture}[scale=0.75]
\begin{axis}[
height=6cm,
width=7.55cm,
legend style={
cells={anchor=east},
at={(0.5,1.55)},anchor=north,
legend plot pos=right
},
xlabel={$p$},
legend columns=3,
ymax=100, ymin=30,
xmin=8, xmax=125,
grid=major,
symbolic x coords={1,8,27,64,125},
cycle multi list={
color list\nextlist
[3 of]mark list
}]]

\addplot coordinates {
(8, 0.973)
(27, 0.962)
(64, 0.942)
(125, 0.933)
}; \addlegendentry{\framework-OpenMPI: 12600}

\addplot coordinates {
(8, 65.0740776911)
(27, 90.2924872028)
(64, 38.1767922454)
(125, 41.295459781)
}; \addlegendentry{6300}

\addplot coordinates {
(8, 90.6946809097)
(27, 82.6421344058)
(64, 44.2629475309)
(125, 32.9899031724)
}; \addlegendentry{4200}

\addplot coordinates {
(8,  96.2440140641)
(27, 92.0529228108)
(64, 92.79976508)
(125, 90.4466111398)
}; \addlegendentry{\framework-MPJ-Express: 12600}

\addplot coordinates {
(8, 94.523006935)
(27, 88.1426660789)
(64, 84.6283571943)
(125, 76.4863733334)
}; \addlegendentry{6300}

\addplot coordinates {
(8, 92.7606189714)
(27, 78.8611217205)
(64, 73.7715792182)
(125, 59.231871605)
}; \addlegendentry{4200}

\addplot coordinates {
(8, 95.6911763854)
(27, 95.0891108662)
(64, 87.0401849801)
(125, 84.4747460585)
}; \addlegendentry{\framework-FastMPJ: 12600}

\addplot coordinates {
(8, 94.3931676398)
(27, 91.1004736654)
(64, 88.1002897971)
(125, 82.2049806855)
}; \addlegendentry{6300}

\addplot coordinates {
(8, 92.9724012065)
(27, 82.6421344058)
(64, 73.7715792182)
(125, 57.9156077916)
 }; \addlegendentry{4200}

\addplot coordinates {
(8, 95.4005741144)
(27, 94.8244272707)
(64, 94.7358041402)
(125, 94.5033305893)
}; \addlegendentry{C-MPI: 12600}

\addplot coordinates {
(8, 96.6590795232)
(27, 94.1695150415)
(64, 93.1560350208)
(125, 92.0731580344)
}; \addlegendentry{6300}

\addplot coordinates {
(8, 96.2024307655)
(27, 92.6104434374)
(64, 91.7341989594)
(125, 81.2331250388)
 }; \addlegendentry{4200}

%
%

\end{axis}
\end{tikzpicture}
\end{minipage}
\caption{Efficiency results for matrix-matrix multiplication (size $n\times n$)
with Grid Abstraction; x-axis: number of cores used; the value for $n$ and the
communication backend employed are given in the legend. Left: results on Carver,
Right: results on {Horseshoe-6}; efficiency is given relative to empirical peak
performance on one core (see text).}
\label{fig:res}
\end{figure}
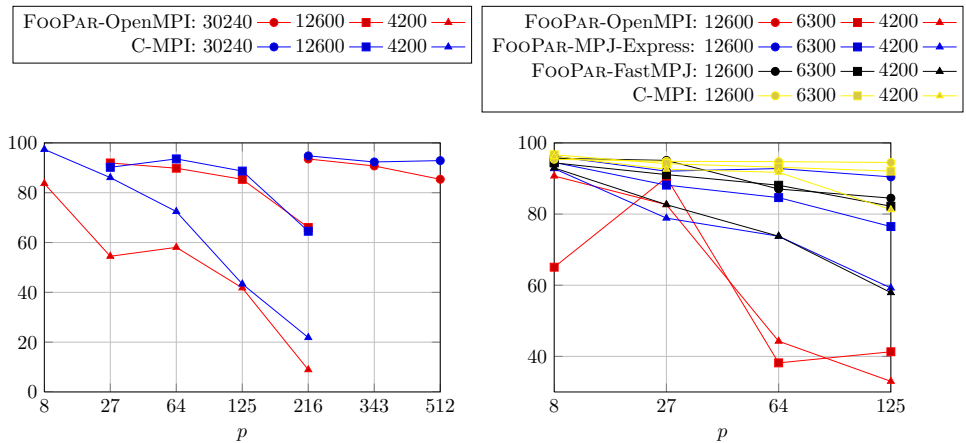

\todo{felix: double-check next paragraph, is it ``still correct''?}
~\\{\noindent \bf Results on Carver:} Efficiencies for different matrix sizes,
$n$, and number of cores, $p$, are given in Fig.\ref{fig:res}. As communication
backend, we used OpenMPI. We note that we improved the Java implementation of
\texttt{MPI\_Reduce} in OpenMPI: the nightly build version implements an
unnecessarily simplistic reduction with $\Theta(p)$ send/receive calls, although
this can be realized with $\Theta(\log p)$ calls. I.e., the unmodified OpenMPI
does {\em not} interface to the native \texttt{MPI\_Reduce} function, and
therefore introduces an unnecessary bottleneck. 

For matrix sizes $n=40000$ and the largest number of cores possible (i.e.
$p=512$) Algorithm~\ref{alg:gridmult} achieves 4.84 TFlop/s, corresponding to
88.8\% efficiency w.r.t. the theoretical peak performance (i.e. 93.7\% of the
empirically achievable peak performance) of Carver. The C-version performs only
slightly better. Note, that the stronger efficiency drop (when compared to
Horseshoe-6 results for smaller matrices) is due to the high performing math
libraries; the absolute performance is still better by a factor of $\approx
2.2$. We conclude that the computation and communication overhead of using
\framework is neglectable for practical purposes. While keeping the advantages
of higher-level constructs, we manage to keep the efficiency very high. This
result is in line with the isoefficiency analysis of
\framework in Section~\ref{sec:mat}.

~\\{\noindent \bf Results on Horseshoe-6:} On Horseshoe-6 we observed that the
different backends lead to rather different efficiencies. When using the
unmodified OpenMPI as a communication backend, a performance drop is seen, as
expected, due to the reasons mentioned above. Also MPJ-Express uses an
unnecessary $\Theta(p)$ reduction (FastMPJ is closed source). However,
if \framework will not be
used in an HPC setting and efficiency is not be the main
objective (like in a heterogeneous system or a cloud environment), the advantages of
``slower'' backends (like running in daemon mode) might pay off. 

\section{Conclusions}
\label{sec:con}
We introduced \framework, a functional and object-oriented framework that
combines two orthogonal scalabilities, namely the scalability as seen from the
perspective of the Scala programming language and the scalability as seen from
the HPC perspective. \framework allows for isoefficiency analyses of algorithms
such that theoretical scalability behavior can be shown. We presented parallel
solutions in \framework for
matrix-matrix multiplication and supported the theoretical
finding with empirical tests that reached close-to-optimal performance w.r.t.
the theoretical peak performance on 512 cores.
\subsubsection*{Acknowledgments:}
We acknowledge the support of the Danish Council for Independent Research, the Innovation Center Denmark, the Lawrence Berkeley National Laboratory, and the Scientific Discovery through Advanced Computing (SciDAC) Outreach Center. We thank Jakob~L. Andersen for supplying a template C-implementation of the DNS algorithm.
\bibliographystyle{abbrv}
\vspace*{-0.1cm}\bibliography{ParaScala-2013}

\begin{thebibliography}{10}

\bibitem{enter}
Scala in the enterprise.
\newblock Ecole Polytechnique Federale de Lausanne (EPFL), \url{
  http://www.scala-lang.org/node/1658}, Accessed 04 May 2013.

\bibitem{jmat}
P.~Abeles.
\newblock Java-{M}atrix-{B}enchmark - a benchmark for computational efficiency,
  memory usage and stability of {J}ava matrix libraries.
\newblock \url{http://code.google.com/p/java-matrix-benchmark/} (accessed
  12.~Feb.~2013).

\bibitem{hiso}
J.~L. Bosque, O.~D. Robles, P.~Toharia, and L.~Pastor.
\newblock H-isoefficiency: Scability metric for heterogenous systems.
\newblock In {\em Proc. of the 10th International Conference of Computational
  and Mathmatical Methods in Science and Engineering (CEMMSE 2010)}, pages
  240--250, 2010.

\bibitem{spmd}
F.~Darema.
\newblock The {S}{P}{M}{D} model : Past, present and future.
\newblock In Y.~Cotronis and J.~Dongarra, editors, {\em Proc. of the 8th
  EuroPVM/MPI Conference}, number 2131 in LNCS, page~1, 2001.

\bibitem{dns81}
E.~Dekel, D.~Nassimi, and S.~Sahni.
\newblock Parallel matrix and graph algorithms.
\newblock {\em SIAM Journal on Computing}, 10(4):657--675, 1981.

\bibitem{gabriel04}
E.~Gabriel, G.~E. Fagg, G.~Bosilca, T.~Angskun, J.~J. Dongarra, J.~M. Squyres,
  V.~Sahay, P.~Kambadur, B.~Barrett, A.~Lumsdaine, R.~H. Castain, D.~J. Daniel,
  R.~L. Graham, and T.~S. Woodall.
\newblock Open {MPI}: Goals, concept, and design of a next generation {MPI}
  implementation.
\newblock In {\em Proc. of the 11th European PVM/MPI Users' Group Meeting},
  pages 97--104, 2004.

\bibitem{Grama93}
A.~Grama, A.~Gupta, and V.~Kumar.
\newblock Isoefficiency: measuring the scalability of parallel algorithms and
  architectures.
\newblock {\em IEEE parallel and distributed technology: systems and
  applications}, 1(3):12--21, 1993.

\bibitem{grama}
A.~Grama, G.~Karypis, V.~Kumar, and A.~Gupta.
\newblock {\em Introduction to Parallel Computing}.
\newblock Pearson, Addison Wesley, 2003.

\bibitem{dns}
A.~Gupta and V.~Kumar.
\newblock Scalability of parallel algorithms for matrix multiplication.
\newblock {\em Proc. of the 22nd International Conference on Parallel
  Processing, ICPP}, 3:115--123, 1993.

\bibitem{comp}
R.~Hundt.
\newblock Loop recognition in {C++/Java/Go/Scala}.
\newblock In {\em Proc. of Scala Days}, 2011.

\bibitem{Hwang98}
K.~Hwang and Z.~Xu.
\newblock {\em Scalable Parallel Computing}.
\newblock McGraw-Hill, New York, 1998.

\bibitem{Kumar87}
V.~Kumar and V.~N. Rao.
\newblock Parallel depth first search. {P}art {II}. {A}nalysis.
\newblock {\em International Journal of Parallel Programming}, 16(6):501--519,
  1987.

\bibitem{eden05}
R.~Loogen, Y.~Ortega-Mall\'{e}n, and R.~{Pe\~{n}a}.
\newblock Parallel functional programming in {E}den.
\newblock {\em Journal of Functional Programming}, 15:431--475, 2005.

\bibitem{scalref}
M.~Odersky.
\newblock The {S}cala language specification, 2011.

\bibitem{fsttcs2009}
M.~Odersky and A.~Moors.
\newblock Fighting bit rot with types (experience report: Scala collections).
\newblock In {\em Proc. of the 29th IARCS Annual Conference on Foundations of
  Software Technology and Theoretical Computer Science (FSTTCS 2009)}, volume~4
  of {\em Leibniz International Proceedings in Informatics}, pages 427--451,
  2009.

\bibitem{pis}
M.~Odersky, L.~Spoon, and B.~Venners.
\newblock {\em Programming in Scala}.
\newblock Artima, 2010.

\bibitem{shafi09}
A.~Shafi and J.~Manzoor.
\newblock Towards efficient shared memory communications in {M}{P}{J} express.
\newblock In {\em Proc. of the 25th IEEE International Symposium on Parallel
  Distributed Processing 2009 (IPDPS)}, pages 1 --7, 2009.

\bibitem{taboada_jos10}
G.~L. Taboada, J.~Touri\~no, and R.~Doallo.
\newblock {F-{M}{P}{J}: Scalable Java Message-passing Communications on
  Parallel Systems}.
\newblock {\em Journal of Supercomputing}, (1):117--140, 2012.

\bibitem{spark10}
M.~Zaharia, N.~M.~M. Chowdhury, M.~Franklin, S.~Shenker, and I.~Stoica.
\newblock Spark: Cluster computing with working sets.
\newblock Technical Report UCB/EECS-2010-53, EECS Department, University of
  California, Berkeley, 2010.

\end{thebibliography}

\end{document}